\documentclass[12pt, letterpaper]{article}
\usepackage{geometry}
\usepackage{amsmath}
\usepackage{graphicx}
\usepackage{float}
\usepackage{siunitx}
\usepackage{subcaption}
\usepackage{amssymb}
\usepackage{cite}
\newgeometry{vmargin={18mm}, hmargin={18mm,18mm}}
\usepackage{mathtools}

\DeclarePairedDelimiterX\braket[2]{\langle}{\rangle}{#1\,\delimsize\vert\,\mathopen{}#2}

\providecommand{\keywords}[1]
{
  \small	
  \textbf{\textit{Keywords---}} #1
}
\begin{document}
\title{Lambert W-Kink Solitons Arising from Higher-Order Nonlinearities of Lipid Membranes}
\author{V. A. Mendoza-Mill\'an$^{1}\footnote{vialezx@gmail.com}$, J. L. Larios-Ferrer$^{2}\footnote{leonel.larios@upenergia.edu.mx}$, J. Samuel Mill\'an$^{2}\footnote{jsmmalo@gmail.com}$,\\ O. Pav\'on-Torres$^{3}\footnote{omar.pavon@cinvestav.mx}$}
\date{%
\small{$^{1}$ Facultad de Ciencias, Universidad Autónoma del Estado de México, Toluca 50200, M\'exico\\
\small{$^{2}$ Universidad Polit\'ecnica de la Energ\'ia, 42820, Tula de Allende, Hidalgo, México\\
\small{$^{3}$ Physics Department, Cinvestav, POB 14-740, 07000 M\'exico City, M\'exico}\\ 
}}}
\maketitle
\begin{abstract}
Accurate modelling of nerve impulse propagation requires accounting for strong higher-order nonlinearities in membrane dynamics, as incorporated in the extended Heimburg–Jackson model. By introducing third- and fourth-order polynomial terms into the membrane density equation, we derive a generalized Duffing-type equation that better captures the complex biophysical states involved in signal transmission. Applying the factorization method, we construct exact travelling wave solutions, including a novel class of Lambert W-Kink-type solitons. These findings provide new analytical insight into the nonlinear electromechanical behaviour of nerve membranes and contribute to the theoretical foundation for understanding pulse propagation in biomembranes.
\end{abstract}

\! \! \! \! \keywords{Lipid membranes, Nerve impulses, Factorization method, Lambert W-Kink solitons.}

\section{Introduction}

The development of neuromorphic devices holds significant promise for advancing our understanding of the human brain and central nervous system by emulating their intricate structures and functions \cite{k1}. Progress in this field has enabled the design of novel technologies with applications ranging from artificial intelligence to neural prosthetics \cite{k00}. In particular, the use of artificially engineered lipid bilayers, memristive components, and synthetic neural networks has become central to both experimental and theoretical research  \cite{k0, k01, k2, k3, k4, k5}. Despite these advances, the intricate behaviour of ion channels and their regulatory mechanisms continues to challenge our understanding of neural signal transmission. This signalling process begins as a localized biochemical event at the axon hillock and propagates along the nerve axon as an electromechanical wave, driven by ion flux across the neuronal membrane \cite{k6, k7}.  A promising theoretical framework for modelling these signals is the thermodynamic soliton theory or Heimburg-Jackson (HJ) model \cite{k8}. This theory describes nerve impulses as nonlinear density waves, arising from the coupling of mechanical, electrical, and thermal processes in the membrane. The model is now widely studied by physicists and mathematicians, especially because the associated density wave equation offers a robust framework for employing quasi-analytical methods \cite{k10, k11, k12, k13, k14, k15, k16, k17, k19}. Although initially met with scepticism, particularly due to the involvement of phase transitions that had not yet been observed experimentally, recent findings have resolved these doubts confirming the presence of distinct sharp phase transitions in nerve membranes during signal conduction \cite{k9}. In fact, the melting transition of the lipid membrane plays a critical role in signal transmission; thus, it is crucial to examine the parameters that influence this phase change, such as pressure and compressibility.

As a consequence, subsequent refinements of the HJ model have introduced additional biophysical parameters to more accurately represent the dynamics of biomembranes observed in experimental studies. These include a viscous term, which models the axoplasmic fluid surrounding the nerve fibre, and an inertial term, implemented as a fourth-order spatial derivative. The latter enhancement, incorporated into the extended HJ model, addresses instabilities that can arise in formulations relying on second-order spatial derivatives, as in the original version \cite{k20, k21, k22, k23}.
In addition to the previously mentioned, recent studies of strong nonlinearities and their impact on modulated waves within the extended HJ model have improved the description of distinct phases of nerve signal transmission, including hyperpolarization, depolarization, overshoot, repolarization, and the refractory period \cite{k18}. Kink soliton solutions effectively characterize these phases, capturing the transitions between excited and relaxed membrane states \cite{k24}. In this context, the primary objective of the present study is to investigate the effects of strong higher-order nonlinearities on the formation of novel soliton structures, such as W-Lambert kink solitons, which represent more suitable candidates for modelling the transmission of nerve signals.

To this end, we begin Section 2 with an overview of the improved HJ model, which incorporates third- and fourth-order nonlinearities. From this formulation, we derive the main governing equation for density waves. It can be shown that the resulting equation assumes the form of a Duffing-type equation, with nonlinear terms expressed as a higher-order polynomial. While quasi-analytical methods can, in principle, be used to study this equation, the combined effects of higher-order nonlinearity and dispersion give rise to a highly intricate and overdetermined system, making such methods insufficient for deriving exact solutions. Consequently, alternative approaches are necessary for obtaining exact travelling wave solutions. In this regard, the factorization method proves to be a powerful tool for addressing polynomial nonlinearities \cite{k25, k26, k27, k28, k29}. Accordingly, Section 3 presents this method and applies it to the density wave equation in nerves, leading to the derivation of Lambert W-Kink-type solitons and other exact travelling wave solutions of neurophysiological interest. Finally, Section 4 offers a comprehensive discussion of the results and their implications.
\section{The mathematical model}
\begin{figure}[ht]
\includegraphics[width=0.5\textwidth]{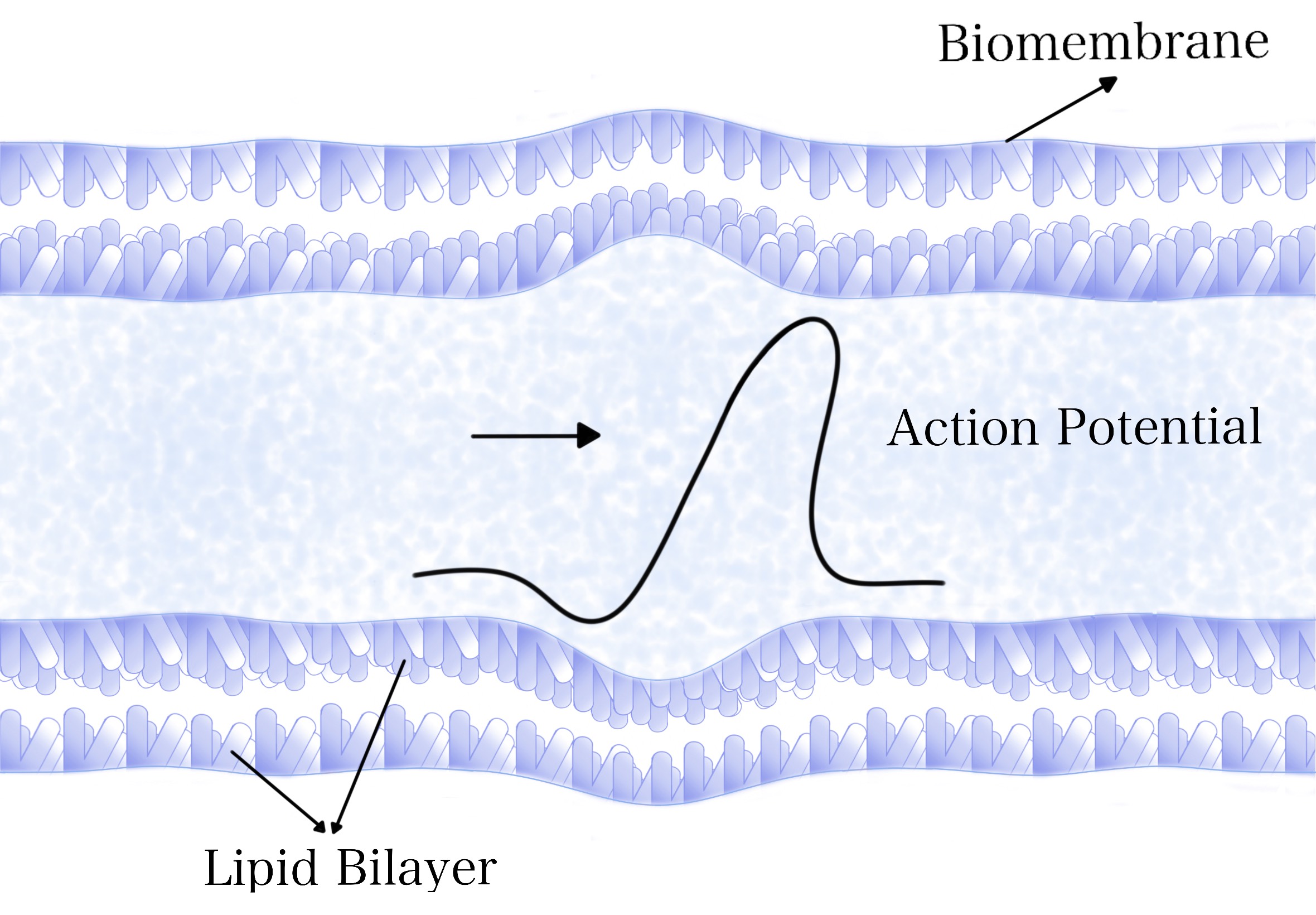}
\centering
\caption{Schematic representation of nerve axon.}\label{ig1}
\end{figure}

Neurons are excitable cells that generate action potentials, which propagate along axons as electromechanical waves. Nerve axons are enclosed by biomembranes composed of phospholipid molecules that spontaneously assemble into bilayers, with hydrophobic tails facing inward and hydrophilic heads outward. As the waves travel, they deform the bilayer, a process governed by the membrane's nonlinear and inelastic mechanical properties, as illustrated in Fig. \ref{ig1}. This behaviour can be described by an extension of the improved HJ model \cite{k18}. 

Thus, let $u=\rho^{A}-\rho_{0}^{A}$ denote the longitudinal density change, defined as the difference between the lateral mass density of the membrane $\rho ^{A}$ and  its empirical equilibrium value $\rho_{0}^{A}$. The governing equation is given by:
\begin{equation}
\dfrac{\partial^{2}u}{\partial {t} ^{2}}=\dfrac{\partial }{\partial x}\left(\left[c_{0}^{2}+\alpha u+\beta u^{2}+\epsilon u^{3}+\lambda u^{4}\right]\dfrac{\partial u}{\partial x}\right)-h_{1} \dfrac{\partial^{4}u}{\partial x ^{4}}+h_{2}\dfrac{\partial^{4}u}{\partial x^{2}\partial t ^{2}}+\mu \dfrac{\partial^{2}}{\partial x^{2}}\left(\dfrac{\partial u}{\partial t}\right). \label{ferrer1}
\end{equation}
Upon close inspection and neglecting the terms proportional to $h_{1}$, $h_{2}$ and $\mu$, we recover the standard form of the wave equation. This reveals that the model relates the propagation velocity $c$ to the lateral compressibility of the circular lipid biomembrane, which is a key assumption of the original HJ model \cite{k8}. The velocity of sound in the fluid phase of the membrane is given by $c_{0}=1/\sqrt{K_{s} ^{A}\rho_{0}^{A}}$, where $K_{s} ^{A}$ is the lateral compressibility. The nonlinear elastic coefficients $\alpha$, $\beta$, $\epsilon$, and $\lambda$ are determined empirically and characterize distinct mechanical effects including lateral compressibility and stretching of the axonal membrane, lipid rarefaction, compositional heterogeneity in lipid–protein interactions, and large-scale deformations arising from compressive and tensile loading. The term associated with $h_{1}$ reflects the elasticity of the biomembrane, while $h_{2}$ accounts for the inertia of the lipid molecules within the membrane. The inclusion of the $h_{2}$ term transforms the HJ equation into a double-dispersion model, which naturally eliminates the instabilities that can arise when only spatial derivatives are present, as in the original formulation. Finally, the term proportional to $\mu$ represents viscous damping due to the surrounding axoplasmic fluid \cite{k24}. 
\subsection{Density wave equation}
In order to study the emergence of novel soliton solutions as response to higher-order nonlinearities, we consider the following re-parametrization of Eq. (\ref{ferrer1}): 
\begin{equation}
y=\dfrac{u}{\rho_{0} ^{A}}, \quad  z=\dfrac{c_{0}x}{\sqrt{h_{1}}}, \quad \text{and} \quad \tilde{t}=\dfrac{c_{0} ^{2}t}{\sqrt{h_{1}}},
\end{equation}
together with
\begin{equation*}
p=\dfrac{\alpha \rho_{0}^{A}}{c_{0}^{2}}; \quad q=\dfrac{\beta (\rho_{0} ^{A})^{2}}{c_{0} ^{2}}; \quad r=\dfrac{\epsilon (\rho_{0} ^{A}) ^{3}}{c_{0} ^{2}}; \quad s=\dfrac{\lambda (\rho_{0} ^{A}) ^{4} }{c_{0}^{2}};
\end{equation*}
\begin{equation}
\gamma=\dfrac{\mu}{\sqrt{h_{1}}} \quad \text{and} \quad \delta=\dfrac{h_{2}c_{0} ^{2}}{h_{1}}.
\end{equation}
To obtain the following dimensionless equation
\begin{equation}
\dfrac{\partial ^{2}y}{\partial \tilde{t} ^{2}}=\dfrac{\partial }{\partial z}\left(\left[1+py+qy^{2}+ry^{3}+sy ^{4}\right]\dfrac{\partial y}{\partial z}\right)-\dfrac{\partial ^{4}y}{\partial z^{4}}+\delta \dfrac{\partial ^{4}y}{\partial z ^{2} \partial \tilde{t}^{2}}+\gamma \dfrac{\partial ^{3}y}{\partial z^{2}\partial \tilde{t}}. \label{badillo1}
\end{equation}
To find travelling wave solutions of Eq. (\ref{badillo1}), we consider $y(\xi)$ with $\xi=kz-v\tilde{t}$, being $k$ and $v$ real-valued constants. After integrate twice the equation with respect to $\xi$ and choosing $C_{1}=C_{2}=0$, we have
\begin{equation}
\dfrac{d ^{2}y}{d\xi ^{2}}+\tilde{\gamma}\dfrac{dy}{d\xi}-\tilde{o}y-\tilde{p}y^{2}-\tilde{q}y ^{3}-\tilde{r}y ^{4}-\tilde{s}y ^{5}=0 \label{ferrer2}
\end{equation}
with the re-parametrized constants 
\begin{align}
\tilde{\gamma}=\dfrac{v \gamma}{k^{2}-\delta v^{2}}; \quad  \tilde{o}=\dfrac{k ^{2}-v^{2}}{ k^{2}(k ^{2}-\delta v ^{2})}; \quad \tilde{p}=\dfrac{p}{2(k ^{2}-\delta v ^{2})};  \nonumber \\ 
\tilde{q}=\dfrac{q}{3(k^{2}-\delta v ^{2})}; \quad \tilde{r}=\dfrac{r}{4(k ^{2}-\delta v ^{2})}; \quad \tilde{s}=\dfrac{s}{5(k^{2}-\delta v ^{2})}. 
\end{align}
It is evident from Eq. (\ref{ferrer2}) that the system belongs to the class of Liénard-type equations. However, in our case, the term associated with the first derivative is constant, which aligns the equation more closely with a Duffing-type oscillator. Specifically, by setting the constants $\tilde{p} = \tilde{r} = \tilde{s} = 0$ in Eq. (\ref{ferrer2}), the equation reduces to the well-known Duffing oscillator \cite{k27}. The density wave equation derived from the extended and improved Heimburg-Jackson model, Eq. (\ref{ferrer1}), thus leads to a nonlinear differential equation with higher-order polynomial terms, as expressed in Eq. (\ref{ferrer2}), making it amenable to the factorization method.

Two primary objectives arise from our investigation. The first is to explore how strong higher-order nonlinearities contribute to the emergence of novel soliton structures, particularly W-Lambert kink solitons. The second is to compare these newly identified solutions with classical kink solitons. The latter are well-established as key structures in multiple biological processes, where they play a significant role in inhibiting neural activity and mediating cell-type-specific responses to high-frequency stimulation \cite{k32}. 

\section{Electromechanical waves in biomembranes and nerve}

\subsection{Kink solitons and the factorization method}
In general, Eq. (\ref{ferrer2}) can be denoted as
\begin{equation}
\dfrac{d ^{2}y}{d\xi ^{2}}+\tilde{\gamma}\dfrac{dy}{d\xi}+f(y)=0. \label{ferrer3}
\end{equation}
As previously mentioned, this corresponds to a nonlinear damped oscillator-type equation, where $\tilde{\gamma}$ serves as a friction coefficient and $f(y)$ is a nonlinear polynomial. In the original work where the method was introduced, a reaction-diffusion equation was associated with Eq. (\ref{ferrer3}), and the same constant $\tilde{\gamma}$ represented the velocity of the travelling wave front solution \cite{k25}. This observation is significant, since commuting $\phi_1$ with $\phi_2$ leads to a modified equation that yields a supersymmetric soliton pair maintaining the same velocity; that is, the corresponding reaction-diffusion equation exhibits an identical propagation velocity.

By factorizing Eq. (\ref{ferrer3}), the equation takes the form
\begin{equation}
\left[\dfrac{d}{d\xi}-\phi_{2}(y)\right]\left[\dfrac{d}{d\xi}-\phi_{1}(y)\right]y=0. \label{ferrer4}
\end{equation}

To enable factorization, similar terms in Eq. (\ref{ferrer4}) are grouped and matched with those in Eq. (\ref{ferrer3}), resulting in the following conditions: 
\begin{subequations}
\begin{equation}
\phi_{1}\phi_{2}=\dfrac{f(y)}{y}; \label{ferrer5}
\end{equation}
\begin{equation}
\phi_{1}+\phi_{2}+\dfrac{d\phi_{1}}{dy}y=-\tilde{\gamma}.\label{ferrer6}
\end{equation} 
\end{subequations}
As a preliminary step towards achieving one of the central goals of this study, we analyze Eq. (\ref{ferrer2}) with $\tilde{r} = \tilde{s} = 0$. This special case reduces to a FitzHugh–Nagumo-type equation, for which kink soliton solutions can be obtained using the factorization method. Accordingly, we arrive to the following equation:
\begin{equation}
\dfrac{d ^{2}y}{d\xi ^{2}}+\tilde{\gamma}\dfrac{dy}{d\xi}-\tilde{o}y+\tilde{p}y^{2}-\tilde{q}y ^{3}=0. \label{ferrer7}
\end{equation}
The case $p < 0$, $q > 0$ reflects a biomembrane above the lipid melting transition, in which higher-order nonlinearities are neglected, i.e., $r = s = 0$ \cite{k030}. As can be seen, Eq. (\ref{ferrer7}) can be factorized as Eq. (\ref{ferrer4}) once we consider 
\begin{equation}
\phi_{1}=\mp\dfrac{1}{\sqrt{2}}\left[ \sqrt{\tilde{q}}y -\sqrt{\dfrac{\tilde{p}^{2}}{4\tilde{q}}-\tilde{o}}-\dfrac{\tilde{p}}{2\sqrt{\tilde{q}}}\right]; \quad \phi_{2}=\pm \sqrt{2}\left[ \sqrt{\tilde{q}}y +\sqrt{\dfrac{\tilde{p}^{2}}{4\tilde{q}}-\tilde{o}}-\dfrac{\tilde{p}}{2\sqrt{\tilde{q}}}\right], \label{add1}
\end{equation}
with
\begin{equation}
\tilde{\gamma}=\pm \dfrac{3}{\sqrt{2}}\sqrt{\dfrac{\tilde{p}^{2}}{4\tilde{q}}-\tilde{o}}\mp \dfrac{\tilde{p}}{\sqrt{8\tilde{q}}}, \label{add2}
\end{equation}
for which the compatibility condition reads as 
\begin{equation}
\left[\dfrac{d}{d\xi}\mp\dfrac{1}{\sqrt{2}}\left( \sqrt{\tilde{q}}y+ \Omega_{1}\right)\right]y=0. \label{k1}
\end{equation}
Thus, the solutions can be written explicitly as:
\begin{equation}
y_{\pm}=\dfrac{2\Omega_{1}}{1\pm \exp\left[\dfrac{\sqrt{q}\Omega_{1}}{\sqrt{2}}(\xi-\xi_{0})-\dfrac{1}{2}\ln \tilde{q}\right]}, \label{AA1}
\end{equation}
and
\begin{equation}
y_{\pm}=\dfrac{2\Omega_{1}}{1\pm \exp\left[-\dfrac{\sqrt{q}\Omega_{1}}{\sqrt{2}}(\xi-\xi_{0})-\dfrac{1}{2}\ln \tilde{q}\right]}, \label{AA2}
\end{equation}
for $\tilde{\gamma}>0$ and $\tilde{\gamma}<0$, respectively. Additionally, we have defined
\begin{equation}
\Omega_{1}=\dfrac{1}{2\sqrt{\tilde{q}}}\left[\sqrt{\dfrac{\tilde{p}^{2}}{4\tilde{q}}-\tilde{o}}+\dfrac{\tilde{p}}{2\sqrt{\tilde{q}}}\right].
\end{equation}
The hyperbolic forms for Eqs. (\ref{AA1}) and (\ref{AA2}) can be easily derived to yield
\begin{equation}
y_{\pm}=\Omega_{1}\left[1-\tanh \left(\pm\dfrac{\sqrt{\tilde{q}}\Omega_{1}}{\sqrt{2}}(\xi-\xi_{0})-\dfrac{1}{4}\ln \tilde{q}\right)\right],\label{L1}
\end{equation} 
\begin{equation}
y_{\pm}=\Omega_{1}\left[1-\coth \left(\pm \dfrac{\sqrt{\tilde{q}}\Omega_{1}}{\sqrt{2}}(\xi-\xi_{0})-\dfrac{1}{4}\ln \tilde{q}\right)\right],\label{c1}
\end{equation}  
with the plus and minus signs corresponding to $\tilde{\gamma} > 0$ and $\tilde{\gamma} < 0$, respectively.

It worth to point it out that the factorization given by Eqs. (\ref{add1}) is not unique. However, since we are dealing with the problem of mechanical waves in lipid membranes and nerves influenced by the axoplasmic fluid, an alternative factorization will lead to impose additional values to the elastic constants, $\tilde{p}$ and $\tilde{q}$, of the membrane to keep the damping coefficient as a constant. An alternative approach to study the effects of axoplasmic fluid in the dynamics of nerve impulses is the multi-scale approach, for which the former equation (\ref{ferrer1}), with $r=s=0$, reduces to a damped nonlinear Schrödinger equation (NLSE). It can be analysed through different perturbative methods, including the quasi-stationary method for a single impulse and the Karpman-Maslov-Solov'ev perturbation theory for orthodromic and antidromic impulses \cite{k24, kk32}.

\begin{figure}
\begin{subfigure}{.5\textwidth}
  \centering
  \includegraphics[width=.93\linewidth]{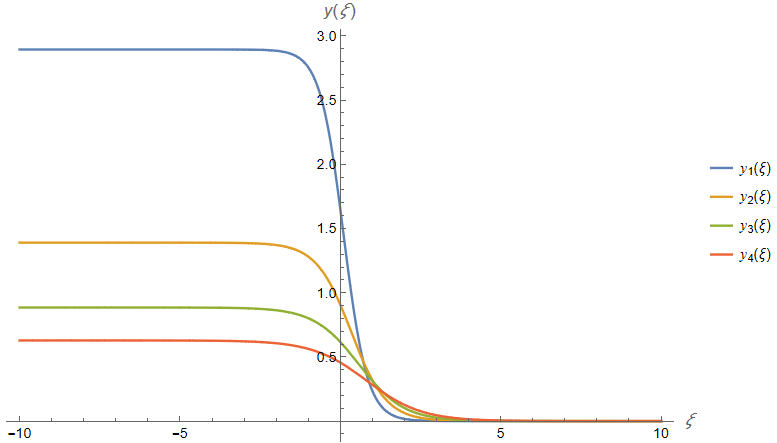}  
  \caption{}
  \label{sub1}
\end{subfigure}
\begin{subfigure}{.5\textwidth}
  \centering
  \includegraphics[width=.93\linewidth]{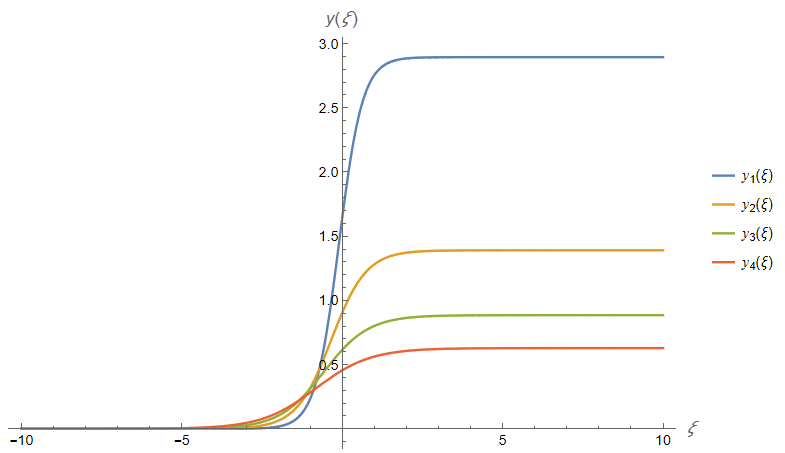}  
  \caption{}
  \label{sub2}
\end{subfigure}
\caption{(a) Kink solitons, and (b) anti-kink solitons, corresponding to Eq. (\ref{L1}) with chosen parameters $k=1$,  $v=0.7$, $\delta= 0.1$, $p=10$ and $q=5$ for $y_{1}$, $q=10$ for $y_{2}$, $q=15$ for $y_{3}$ and $q=20$ for $y_{4}$.}
\label{fig2}
\end{figure}

\begin{figure}
\begin{subfigure}{.5\textwidth}
  \centering
  \includegraphics[width=.98\linewidth]{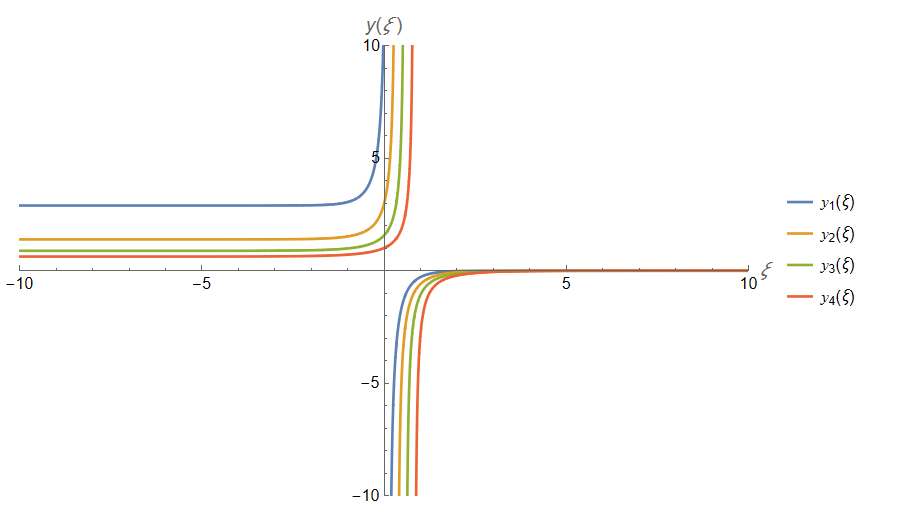}  
  \caption{}
  \label{subbb1}
\end{subfigure}
\begin{subfigure}{.5\textwidth}
  \centering
  \includegraphics[width=.98\linewidth]{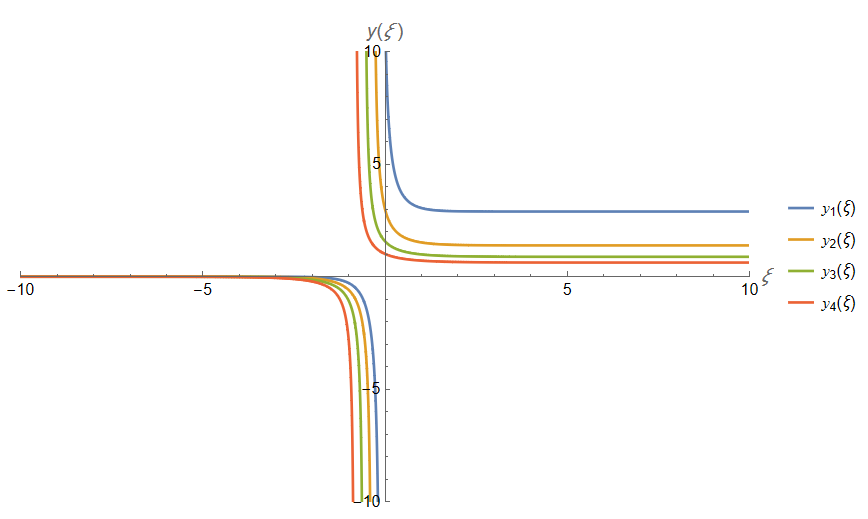}  
  \caption{}
  \label{subbb2}
\end{subfigure}
\caption{(a) $y_{+}(\xi)$ and (b) $y_{-}(\xi)$ corresponding to Eq. (\ref{c1}) with chosen parameters $k=1$,  $v=0.7$, $\delta= 0.1$, $p=10$ and $q=5$ for $y_{1}$, $q=10$ for $y_{2}$, $q=15$ for $y_{3}$ and $q=20$ for $y_{4}$.}
\label{fig4}
\end{figure}

The form of Eq. (\ref{L1}) represents a kink and anti-kink soliton, respectively (depicted in Fig. \ref{sub1} and Fig. \ref{sub2}), which are physically associated with a shock wave of a freezing transition, a travelling disruption that changes the displacement of the membrane in a step-like manner from fluid to gel state, respectively\cite{kkk32}. A plausible physical interpretation of Eqs. (\ref{c1}), can be provided within the framework of the Poisson–Nernst–Planck (PNP) equations, which describe ion transport in the dilute solution limit of electrolyte systems \cite{kk34}. To further explore this interpretation, we consider the formation of diffuse charge layers of finite thickness near the membrane–solution interface. The curves corresponding to each branch of Eq. (\ref{c1}), depicted in Fig. \ref{fig4}, can be associated with the accumulation of pumped charges that result in a positively charged diffuse layer localized near the membrane–fluid interface. The net effect of cation removal from the first domain is to induce an effective negative surface charge on the system formed by the second domain and the intervening insulating membrane \cite{kk35}.

In general, with a suitable choice of parameters, almost any quasi-analytical method can yield travelling wave solutions similar to those obtained via the factorization method for Eq. (\ref{ferrer7}). This is illustrated in Appendix A, where we compare the solutions $y_{\pm}(\xi)$, for both $\tilde{\gamma}>0$ and $\tilde{\gamma}<0$, with those derived using the basic $G'/G$ expansion method.  Such methods are particularly effective for equations where the higher-order dispersion term can be balanced by the nonlinear term, leading to a significant simplification such as Eq. (\ref{ferrer7}). However, when higher-order polynomial nonlinearities are introduced, this balancing becomes either infeasible or the yielding algebraic system is overly intricate. In such scenarios, the factorization method becomes indispensable for capturing the effects of these nonlinearities and exploring their physical implications.
With regard to the nonlinear elastic constants and the density wave solutions of Eq. (\ref{ferrer7}), several well-characterized special cases arise, particularly when $h_{2}=\mu=0$. For $p>0$, $q>0$, and $r=s=0$, solitary waves with negative amplitudes can emerge \cite{k031}. If $p>0$ and $r=s=0$, a single stable solution exists, whereas in the case where $q<0$ and $r=s=0$, two stable solutions of different amplitude may occur; in this scenario, the polarity of the solution is determined by $p$ \cite{k032}. More generally, for the simplified case where $r=s=h_{2}=\mu=0$, Eq. (\ref{ferrer1}) can be integrated twice to yield a Jacobi elliptic equation, which admits a family of doubly periodic solutions expressed in terms of Weierstrass elliptic and Jacobian elliptic functions \cite{kk30} (see Appendix A in \cite{k30} for the complete set of solutions). These solutions depend on the specific values of $p$ and $q$, and are particularly relevant for describing the emergence of periodic soliton trains emergence in the nerve \cite{k31}. A similar approach can be employed using the factorization method to derive alternative travelling wave profiles. Nevertheless, this lies beyond the primary scope of the present work, which focuses on mechanical waves in biomembranes and nerves. 

\subsection{Lambert W-Kink-type solitons}
To analyse the impact of higher-order polynomial nonlinearities in lipid membranes and the consequent emergence of novel soliton-like solutions with potential biophysical relevance, we need to consider an extension of the HJ model. This generalized framework admits new types of solutions with distinctive features, such as Lambert W kink-like solitons. Although both solutions appear closely related, the Lambert W function exhibits asymmetry properties that lead to a fundamentally different physical interpretation. Therefore, similar to the classical case of the preceding section, it is necessary to incorporate the physically relevant parameter regime for biomembranes, specifically $p<0$ and $q>0$ in Eq. (\ref{ferrer2}), which consequently leads to the following equation
\begin{equation}
\dfrac{d ^{2}y}{d\xi ^{2}}+\tilde{\gamma}\dfrac{dy}{d\xi}-\tilde{o}y+\tilde{p}y^{2}-\tilde{q}y ^{3}-\tilde{r}y ^{4}-\tilde{s}y ^{5}=0.\label{di1}
\end{equation}
Closed-form solutions of a generalized version of Eq. (\ref{di1}) have been previously analysed for the special case of $\tilde{\gamma}=0$ using the factorization method \cite{k27}. However, only a limited number of cases yield implicit solutions containing two integration constants. In contrast, the case under consideration with $\tilde{\gamma}\neq 0$ enables to obtain additional information regarding the biomembrane parameters $\tilde{r}$ and $\tilde{s}$. However, the admissible soliton profiles impose strong constraints on these parameters. Although the parameters $\tilde{r}$ and $\tilde{s}$ must be determined experimentally for each specific membrane, the application of the factorization method offers valuable insight into their analytically feasible domain. This, in turn, facilitates the calibration of complementary experimental parameters to ensure consistency with theoretical predictions.

Extending the analysis from previous analysis, we examine Eq. (\ref{di1}), for which Eq. (\ref{ferrer3}) assumes the following form 
\begin{equation}
\left(-\tilde{s}y ^{4}-\tilde{r}y ^{3}-\tilde{q}y ^{2}+\tilde{p}y-\tilde{o}\right)y=\phi_{1}\phi_{2}y, \label{fushim1}
\end{equation}
consequently, yielding to 
\begin{equation}
\phi_{1}=\pm \sqrt{\dfrac{\tilde{s}}{3}}\left(y^{2}+Ey-\tilde{o}\right); \quad \phi_{2}=\pm \sqrt{\dfrac{3}{\tilde{s}}}\left(-\tilde{s}y^{2}+By+1\right)\label{fushim2}
\end{equation}
with 
\begin{equation}
B=-\left(\tilde{s}\tilde{o}+\tilde{q}+1\right)E^{-1},
\end{equation}
and
\begin{equation}
E=\dfrac{\tilde{o}\tilde{r}-\tilde{p}}{\tilde{o}\tilde{s}-1}.
\end{equation}
In order to obtain the $\tilde{\gamma}$, again we replace Eqs. (\ref{fushim2}) in Eq. (\ref{ferrer6}), obtaining 
\begin{equation}
\tilde{\gamma}=\pm \dfrac{\tilde{s}\tilde{o}-3}{\sqrt{3\tilde{s}}},\label{carm1}
\end{equation}
with the values of $\tilde{s}$ provided by
\begin{equation}
\tilde{s}^{3}+a_{2}\tilde{s} ^{2}+a_{1}\tilde{s}+a_{0}=0, \label{cardano}
\end{equation}
for which
\begin{equation}
\tilde{s}=\sqrt[3]{-\dfrac{Q}{2}+\sqrt{\Delta}}+\sqrt[3]{-\dfrac{Q}{2}- \sqrt{\Delta}}-\dfrac{a_{2}}{3},\label{milesahead1}
\end{equation}
where
\begin{equation}
\Delta=\dfrac{Q ^{2}}{4}+\dfrac{P^{3}}{27}; \label{milesahead2}
\end{equation}
\begin{subequations}
\begin{equation}
P=a_{1}-\dfrac{a_{2}^{2}}{3}; \label{milesahead3}
\end{equation}
\begin{equation}
Q=\dfrac{2}{27}a_{2}^{3}- \dfrac{1}{3}a_{2}a_{1}+a_{0}. \label{milesahead4}
\end{equation}
\end{subequations}
Along with
\begin{subequations}
\begin{equation}
a_{0}=\dfrac{1}{\tilde{o}^{3}}\left(\tilde{q}+1\right),\label{shack1}
\end{equation}
\begin{equation}
a_{1}=-\dfrac{1}{3\tilde{o}^{3}}\left[{2(\tilde{p}-\tilde{o}\tilde{r})^{2}}+{3\tilde{o}\left(2\tilde{q}+1\right)}\right],\label{shack2}
\end{equation}
\begin{equation}
a_{2}=\dfrac{1}{\tilde{o}}\left(\tilde{q}-1\right).\label{shack3}
\end{equation}
\end{subequations}
Since $\tilde{s}$ is a physical parameter which characterizes to the membrane, it is expected to be real-valued and positive definite, in accordance with $\tilde{\gamma}$, provided by Eq. (\ref{carm1}). To ensure this condition is satisfied, we must take into account the nature of the roots of the associated with the cubic equation Eq. (\ref{cardano}), through the sign of its discriminant, $\Delta$. Thus, we have that: if $\Delta >0$, the equation yields one real root and two complex conjugate roots;  if $\Delta=0$, all roots are real, with at least two equal, and if $\Delta<0$; all roots are real and distinct. Thus, for the special case of $\tilde{s}$, we have  
\begin{subequations}
\begin{equation}
\sqrt[3]{-\dfrac{Q}{2}+\sqrt{\Delta}}+\sqrt[3]{-\dfrac{Q}{2}- \sqrt{\Delta}}>\dfrac{a_{2}}{3} \quad (\text{if} \, \, \Delta>0),
\end{equation}
\begin{equation}
2\sqrt[3]{-\dfrac{Q}{2}}>\dfrac{a_{2}}{3} \quad (\text{if} \, \, \Delta=0),
\end{equation}
\begin{equation}
2\sqrt{-\dfrac{P}{2}}\cos \left(\dfrac{\theta}{3}\right)>\dfrac{a_{2}}{3} \quad (\text{if} \, \, \Delta<0),\label{juanga1}
\end{equation}
\end{subequations}
for the last equation (\ref{juanga1}), we have defined 
\begin{equation}
\cos \theta=-\dfrac{Q}{2}\left(-\dfrac{3}{P}\right)^{3/2}.
\end{equation}
In order to find closed-form solutions to Eq. (\ref{fushim1}), the compatibility condition reads as 
\begin{equation}
\left[\dfrac{d}{d\xi}\pm \sqrt{\dfrac{\tilde{s}}{3}}\left(y^{2}+Ey-\tilde{o}\right) \right]y=0,\label{krafty1}
\end{equation} 
and, after some calculations, we obtain
\begin{equation}
y_{\pm}=\bar{\alpha} \left[1-\dfrac{1}{1+ W\left[\varphi(\xi)\right]}\right],\label{u1}
\end{equation}
\begin{equation}
\varphi(\xi)=\exp\left(\mp \bar{\alpha}^{2}\sqrt{\dfrac{\tilde{s}}{3}}\xi- 1\right)\label{u2}
\end{equation}
for $\tilde{\gamma}>0$ and $\tilde{\gamma}<0$, respectively. Here, $W[\varphi(\xi)]$ in Eq. (\ref{u1}) denotes the Lambert W function, defined as the inverse function of $f(W)=We^{W}$, and $\bar{\alpha}$ is a real parameter given by 
\begin{equation}
\bar{\alpha}=\sqrt[3]{-\dfrac{\bar{Q}}{2}+\sqrt{\bar{\Delta}}}+\sqrt[3]{-\dfrac{\bar{Q}}{2}- \sqrt{\bar{\Delta}}}-\dfrac{\bar{a}_{2}}{3},
\end{equation}
where
\begin{equation}
\bar{\Delta}=\dfrac{\bar{Q} ^{2}}{4}+\dfrac{\bar{P}^{3}}{27}; 
\end{equation}
\begin{subequations}
\begin{equation}
\bar{P}=\bar{a}_{1}-\dfrac{\bar{a}_{2}^{2}}{3}; 
\end{equation}
\begin{equation}
\bar{Q}=\dfrac{2}{27}\bar{a}_{2}^{3}- \dfrac{1}{3}\bar{a}_{2}\bar{a}_{1}+\bar{a}_{0}; 
\end{equation}
\end{subequations}
and
\begin{equation}
\bar{a}_{0}=- \dfrac{\tilde{p}}{2\tilde{s}}; \quad \bar{a}_{1}=\dfrac{1}{\tilde{s}}; \quad \bar{a}_{2}=- \dfrac{\tilde{r}}{2\bar{s}};
\end{equation}
for which a similar analysis to $\tilde{s}$ in Eq. (\ref{milesahead1}) was carried out for $\bar{\alpha}$, a real parameter, either negative or positive. Thus, again we must be carefully with discriminant $\bar{\Delta}$. However, in this case, we no longer require the condition that $\bar{\alpha}$ be positive definite. 

\begin{figure}
\includegraphics[width=0.5\textwidth]{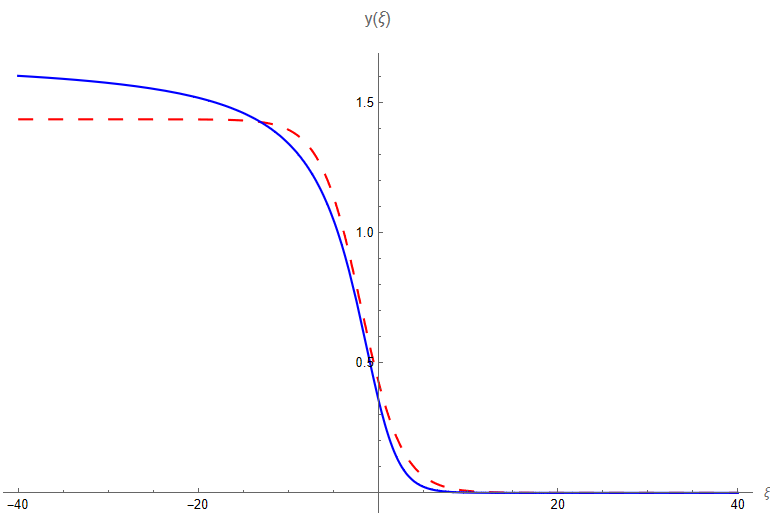}
\centering
\caption{Comparison between the Lambert W-kink soliton (solid line) and the standard kink (dashed line).}\label{igg1}
\end{figure}
\begin{figure}
\begin{subfigure}{.5\textwidth}
  \centering
  \includegraphics[width=.99\linewidth]{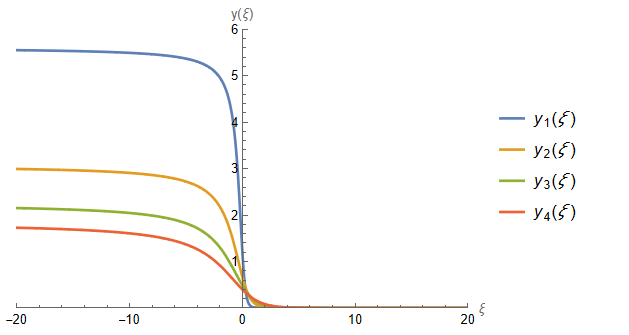}  
  \caption{}
  \label{sub3}
\end{subfigure}
\begin{subfigure}{.5\textwidth}
  \centering
  \includegraphics[width=.99\linewidth]{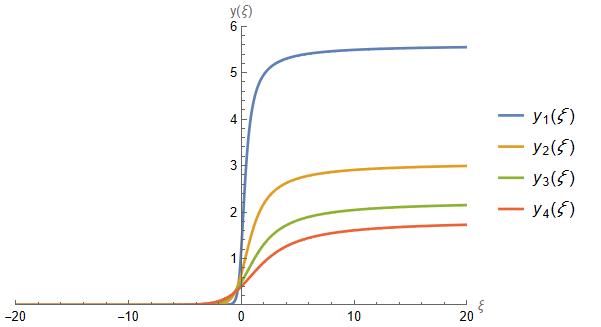}  
  \caption{}
  \label{sub4}
\end{subfigure}
\begin{subfigure}{.5\textwidth}
  \centering
  \includegraphics[width=.99\linewidth]{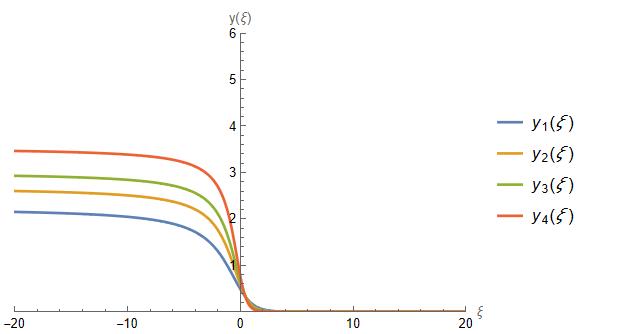}  
  \caption{}
  \label{sub5}
\end{subfigure}
\begin{subfigure}{.5\textwidth}
  \centering
  \includegraphics[width=.99\linewidth]{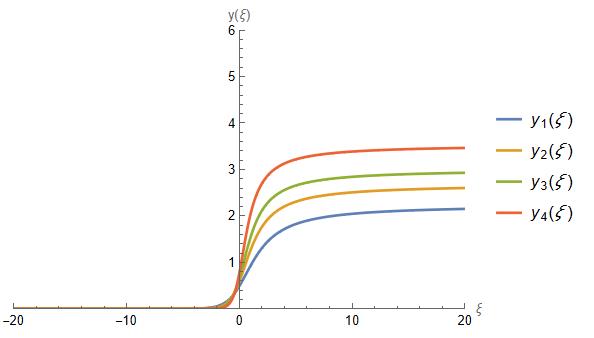}  
  \caption{}
  \label{sub6}
\end{subfigure}
\caption{(a) and (b) Lambert W-kink-type solitons, corresponding $y_{+}$ provided by Eq. (\ref{u1}) with Eq. (\ref{u2}) with chosen parameters  $k=0.5$,  $v=0.6$, $\delta=2.1$, $p=4.5$, $r=6$ and $q=2$ for $y_{1}$, $q=2.5$ for $y_{2}$, $q=3$ for $y_{3}$ and $q=3.5$ for $y_{4}$. (c) and (d) Lambert W-kink-type solitons, corresponding to Eq. (\ref{u1}) with Eq. (\ref{u2}) with chosen parameters  $k=0.5$,  $v=0.6$, $\delta=2.1$, $q=4.5$, $r=6$ and $p=4.5$ for $y_{1}$, $p=4.8$ for $y_{2}$, $p=5$ for $y_{3}$ and $p=5.3$ for $y_{4}$.}
\label{fig3}
\end{figure}	
The Lambert W-kink-type solution ($\ref{u1}$) was studied in the framework of the $\phi^{6}$-theory and contrasted to the $\phi^{4}$-theory of high-energy physics \cite{k34}\footnote{The factorization performed for Eq. (\ref{fushim1}) is a particular of
\begin{equation*}
-\tilde{s}Y ^{4}-\tilde{r}Y^{3}-\tilde{q}Y^{2}+\tilde{p}Y-\tilde{o}=\left(DY ^{2}+EY+F\right)\left(AY ^{2}+BY+C\right)
\end{equation*}
which, lead to the overdetermined system: 
\begin{subequations}
\begin{equation*}
AD=-\tilde{s};\label{au1}
\end{equation*}
\begin{equation*}
AE+BD=-\tilde{r};\label{au2}
\end{equation*}
\begin{equation*}
AF+BE+CD=-\tilde{q}
\end{equation*}
\begin{equation*}
BF+CE=\tilde{p};\label{au3}
\end{equation*}
\begin{equation*}
CF=-\tilde{o}.\label{au4}
\end{equation*}
\end{subequations}
As a particular case $C=D=1$ was considered which lead to Eqs. (\ref{fushim2}). However, more general factorization can be obtained once we solve the overdetermined system of algebraic equations with symbolic computation.}. For our case, we can see that higher-order polynomial nonlinearities produce transcendental behaviour. Therefore, the biological phenomena cannot be well described by the standard kink solitons, but with the Lambert W-kink solitons as a response (see Fig. \ref{igg1}). Thus, the W-kink solitons arise as a response to the effects of higher-order polynomial nonlinearities to describe the transition between two membrane states. Since one of the mathematical properties of this kind of soliton is their asymmetry, it could be interpreted biologically as nonlinear recovery or hysteresis in membrane behaviour. On the other side, the branches in the behaviour of the Lambert W function, depicted in Fig. \ref{fig3}, could be interpreted as a possible bi-stability or threshold excitation. 

\begin{figure}
\includegraphics[width=0.7\textwidth]{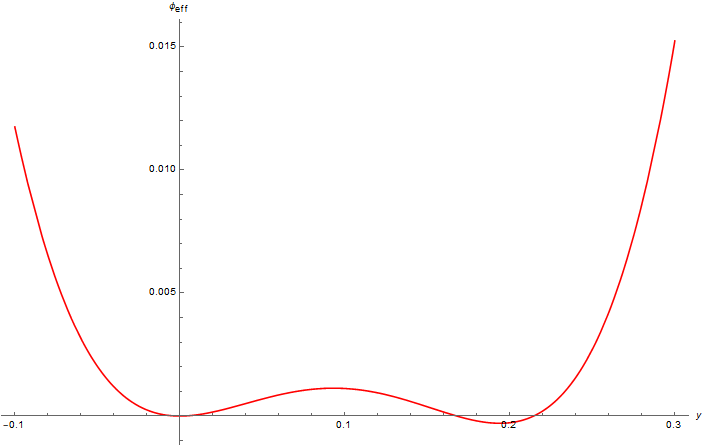}
\centering
\caption{Effective potential $\phi_{eff}(y)$ with $p=5$, $q=15.3$, $r=13.1$, $s=-12.8$ and $\delta=1.4$.}\label{ig4}
\end{figure}

Finally, to identify the regime in which solitary wave solutions may exist for Eq. (\ref{ferrer2}) with $\tilde{\gamma} = 0$, we can adopt a mechanical analogy \cite{k21}. By performing two successive integrations of Eq. (\ref{ferrer2}), we readily obtain:
\begin{equation}
(k^{2}-\delta v^{2})(y')^{2}=\dfrac{k^{2}-v^{2}}{k^{2}}y^{2}+\dfrac{p}{3}y^{3}+\dfrac{q}{6}y^{4}+\dfrac{r}{10}y ^{5}+\dfrac{s}{15}y ^{6}, \label{yogurt1}
\end{equation}
where the right side of Eq. (\ref{yogurt1}) can be seen as a pseudo-potential. In the present case of lipid membranes, it is required that $p<0$, $q>0$, $\tilde{s}>0$. Under these conditions,  solitary wave solutions are possible when the effective potential $\Phi_{eff}(y)$ is positive and exhibits a local minimum at $\phi_{eff}(y)=0$, accompanied by at least one adjacent local maximum, as illustrated in Fig. \ref{ig4}.

As a final remark, and similarly to the case without higher-order nonlinearities discussed in the previous section, the direct application of the multi-scale method reduces Eq. (\ref{ferrer1}) to a damped cubic–quintic NLSE. This reduced form is well-suited for analysis through modulational instability and numerical simulations. To explore the adiabatic evolution of solitons in lipid membranes from an alternative perspective, a quasi-stationary approach can be employed for both bright and dark quasi-solitons. However, the first-order perturbative solutions for bright quasi-solitons in the damped cubic–quintic NLSE are significantly more intricate than those derived from the standard damped NLSE. In contrast, a corresponding analysis for dark quasi-solitons remains an open and compelling problem, particularly in evaluating the impact of higher-order nonlinearities on their adiabatic dynamics \cite{k35, k36}.
   
\section{Conclusion}
The elegance of the factorization method lies in its simplicity, though this very simplicity is sometimes underestimated. In this work, we have shown that the method can be systematically applied to an extended nerve model incorporating higher-order nonlinearities, enabling the construction of novel soliton solutions, such as Lambert W–Kink-type solitons emerging from third- and fourth-order polynomial terms. 
Beyond solution construction, the method provides valuable analytical insight into the physical parameters governing these nonlinearities, an advantage not easily achieved using quasi-analytical methods. In such cases, the resulting overdetermined algebraic systems become highly intricate, and balancing procedures are far from straightforward. Although our primary focus has been on W–Kink-type solitons, the extended model admits a wider family of solutions when the polynomial order is further increased, with potential implications for physiology and biophysics.  A more comprehensive exploration of these possibilities includes generalizations beyond biomembranes of the generalized Boussinesq equation with cubic and quartic nonlinearities, something that we will present in future works. 
Among the most compelling outcomes of this analysis is the emergence of supersymmetric soliton pairs, solutions with identical wavefront velocities but governed by distinct equations. While the role of supersymmetry in biological systems remains largely unexplored, our results suggest that it could offer valuable insights into the mechanisms underlying nerve signal transmission. In the extended model, applying supersymmetric analogies leads to two distinct scenarios: either the polynomial nonlinearities are modified, resulting in a new equation, or the damping term becomes a function of the travelling wave variable rather than a constant. Both cases are physiologically significant, especially in the context of nerve impulse propagation and lipid membrane dynamics.The additional terms arising in the first case can be interpreted as an external stimulus acting on the lipid membrane, modifying its potential and, consequently, altering the original soliton-like solution. In contrast, the second case corresponds to a generalized perturbation term, which accounts not only for damping effects but also for any form of mechanical stress affecting the membrane. Exploring these two scenarios deepens our understanding of one of the most intricate processes in biology: the propagation of nerve impulses through lipid membranes.

\section*{Appendix A}
As it will be shown similar exact travelling wave solutions of Eq. (\ref{badillo1}) can be obtained through the basic $G'/G$-expansion method \cite{k37}. The main algorithm of this quasi-analytical method can be summarized as follows:

\begin{enumerate}
\item As an initial step, we consider travelling wave solutions of the form $\xi=kx-vt$ for the nonlinear differential equation represented by $P(u, u_{x}, u_{t}, ..., {x}, t)=0$, thereby reducing it to an ordinary differential equation in terms of the variable $\xi$ 
\begin{equation}
E(U, U_{\xi}, U_{\xi\xi}, ..., \xi)=0. \label{badi1}
\end{equation}
\item In the following, we express the solution $U(\xi)$ as a finite series composed of logarithmic derivatives
\begin{equation}
U(\xi)=\sum^{N}_{k=0}b_{k}\left(\dfrac{G'(\xi)}{G(\xi)}\right)^{k}, \quad G'(\xi)=\dfrac{dG(\xi)}{d\xi},\label{badi2}
\end{equation}
where $b_{k}$ denote parameters to be determined, and $G(\xi)$ satisfies the damped oscillator equation:
\begin{equation}
G''+\tilde{\lambda} G'+\tilde{\mu} G=0, \label{badi3}
\end{equation}
where $\tilde{\lambda}$ and $\tilde{\mu}$ are constants to be determined. 
\item To determine the integer $N$ in the logarithmic derivative series (\ref{badi2}), a homogeneous balance must be performed between the highest-order nonlinear term and the highest-order dispersive term in Eq. (\ref{badi1}).	
\item Subsequently, the proposed solution in Eq. (\ref{badi2}) is substituted into Eq. (\ref{badi1}), leading to a system of algebraic equations. The resulting system enables the determination of the unknown parameters $b_{k}$, $\tilde{\lambda}$, and $\tilde{\mu}$.  
\item Finally, by substituting the previously determined parameters and the solutions of Eq. (\ref{badi3}), the general solution to Eq. (\ref{badi1}) is obtained. 
\end{enumerate}

To implement the main algorithm summarized above, we begin with Eq. (\ref{ferrer7}), the governing equation obtained after applying the traveling wave ansatz from Step 1 to the original model. Next, to express the solution as indicated in Step 2, we apply the homogeneous balance procedure outlined in Step 3, which yields $N = 1$. Consequently, the solution given in Eq. (\ref{badi2}) reduces to the form:
\begin{equation}
y(\xi)=b_{0}+b_{1}\left(\dfrac{G'(\xi)}{G(\xi)}\right).
\end{equation}
By substituting it into Eq. (\ref{ferrer7}), along with Eq. (\ref{badi3}), we obtain the following system of algebraic equations:
\begin{subequations}
\begin{equation}
2b_{1}-\tilde{q}b_{1}^{3}= 0;
\end{equation}
\begin{equation}
3b_{1}\tilde{\lambda} - \tilde{\gamma}b_{1} + \tilde{p} b_{1}^{2} - 3\tilde{q} b_{1}^{2} b_{0}= 0      ;
\end{equation}
\begin{equation}
b_{1} (2 \tilde{\mu} + \tilde{\lambda}^{2}) - \tilde {\gamma}  b_{1} \tilde{\lambda} - \tilde{o} b_{1} +2 \tilde{p} b_{1} b_{0} - 3 \tilde{q} b_{1} b_{0}^{2} = 0;
\end{equation}
\begin{equation}
b_{1} \tilde{\lambda} \tilde{\mu} - \tilde {\gamma}  b_{1} \tilde{\mu} - \tilde{o} b_{0} + \tilde{p} b_{0}^{2} - \tilde{q} b_{0}^{3}= 0.
\end{equation}
\end{subequations}
Upon solving the system of equations, we obtain:
\begin{equation}
b_{0} = \frac{1}{ 3 \tilde{q} b_{1}}(3\tilde{\lambda} - \tilde{\gamma}) + \frac{\tilde{p}}{3 \tilde{q}}; \quad b_{1}= \pm \sqrt{ \frac{2}{\tilde{q}}}.
\end{equation}
The solutions of Eq. (\ref{badi3}) are determined by the characteristic equation:
\begin{equation}
m = \frac{-\tilde{\lambda} \pm \sqrt{\tilde{\lambda}^{2}-4\tilde{\mu}}}{2}, \label{tom1}
\end{equation}
which depends on the values of $\tilde{D}=\tilde{\lambda}^{2}-4\tilde{\mu}$.
\subsubsection*{Case A.} 
If $\tilde{D}=0$, then $m_{1}=m_{2}=-\dfrac{\tilde{\lambda}}{2}$, and the corresponding solution for $y(\xi)$ is
\begin{equation}
y(\xi) = b_{0}+b_{1}\left(-\frac{\tilde{\lambda}}{2}+ \frac{c_{2}}{c_{1}+(\xi- \xi _{1}) c_{2}}\right),
\end{equation}
where $c_{1}$ and $c_{2}$ are arbitrary integration constants.
\subsubsection*{Case B.} 
If $\tilde{D}>0$, then $m_{1}=\dfrac{-\tilde{\lambda}+\sqrt{\tilde{D}}}{2}$ and $m_{2}=\dfrac{-\tilde{\lambda}-\sqrt{\tilde{D}}}{2}$, and the corresponding solutions are 
\begin{equation}
y(\xi)=b_{0}+ b_{1}\left(-\frac{\tilde{\lambda}}{2}+\frac{\sqrt{\tilde{D}}}{2} \tanh\left(\frac{\sqrt{\tilde{D}}(\xi- \xi _{1})}{2}\right)\right), \label{car1}
\end{equation}
for $c_{1}=c_{2}=1$, and
\begin{equation}
y(\xi)=b_{0}+ b_{1}\left(-\frac{\tilde{\lambda}}{2}+\frac{\sqrt{\tilde{D}}}{2} \coth\left(\frac{\sqrt{\tilde{D}}(\xi- \xi _{1})}{2}\right)\right) \label{car2}
\end{equation}
for $c_{1}=1$ and $c_{2}=-1$.

It is readily observed that Eqs. (\ref{car1}) and (\ref{car2}) bear structural similarity to Eqs. (\ref{L1}) and (\ref{c1}). A direct comparison allows us to identify the conditions under which both sets of solutions coincide, which turns out to be
\begin{subequations}
\begin{equation}
b_{0}=\Omega_{1}+\dfrac{\tilde{\lambda}}{2}b_{1},
\end{equation}
\begin{equation}
b_{1}=-\dfrac{2\Omega_{1}}{\sqrt{\tilde{D}}}; \quad {\tilde{D}}={2\tilde{q}}\Omega_{1} ^{2}
\end{equation}
\begin{equation}
\xi_{1}=\xi_{0}+\dfrac{1}{\sqrt{8\tilde{q}}\Omega_{1}}\ln \tilde{q}.
\end{equation}
\end{subequations}
\subsubsection*{Case C.}
If $\tilde{D}<0$, then $m_{1}=\dfrac{-\tilde{\lambda}+\sqrt{\tilde{D}}i}{2}$ and $m_{2}=\dfrac{-\tilde{\lambda}-\sqrt{\tilde{D}}i}{2}$, and the solutions for $y(\xi)$ are
\begin{equation}
y(\xi)=-\dfrac{\lambda}{2}+\dfrac{\sqrt{\tilde{D}}}{2} \frac{ \cos(\sqrt{\tilde{D}} (\xi- \xi _{1}))}{1+ \sin(\sqrt{\tilde{D}} (\xi- \xi _{1}))} ,
\end{equation} 
with $c_{1}=c_{2}=1$. 

It is clear that the algorithm’s structure closely resembles that of many quasi-analytical methods. For example, the tanh-expansion method can be shown to be equivalent to the basic $G'/G$-expansion method \cite{k38}, which itself is a specific case of the Jacobi elliptic function expansion method. This, in turn, is a special case of the generalized $F$-expansion method, and so on \cite{k39}.


\medskip

\noindent \textbf{Data Availability Statement} Data sharing not applicable to this article as no datasets were generated or analyzed during the current study.


\begin{thebibliography}{9}
\bibitem{k1} Xinyi Qi and Giovanni Zocchi. Kink propagation in the Articial Axon. EPL, 137 (2022) 12005. 
\bibitem{k00} Qian Y, Alhaskawi A, Dong Y, Ni J, Abdalbary S and Lu H (2024) Transforming medicine: artificial intelligence integration in the peripheral nervous system. Front. Neurol. 15:1332048.
\bibitem{k0} Alquran, M., Sulaiman, T.A. \& Yusuf, A. Kink-soliton, singular-kink-soliton and singular-periodic solutions for a new two-mode version of the Burger–Huxley model: applications in nerve fibers and liquid crystals. Opt Quant Electron 53, 227 (2021).
\bibitem{k01} Lin, Z., Deng, J., Chen, Z. et al. An efficient reservoir computing system based on 2D mask processing and dynamic memristor. Nonlinear Dyn (2025).
\bibitem{k2} Quan Xu, Yujian Fang, Huagan Wu, Han Bao, Ning Wang. Firing patterns and fast–slow dynamics in an N-type LAM-based FitzHugh–Nagumo circuit. Chaos, Solitons and Fractals 187 (2024) 115376.  
\bibitem{k3} Binchi Wang, Ya Wang, Xiaofeng Zhang, Zhigang, Zhu. A memristive neuron with nonlinear membranes and network patterns. Physics Letters A 540 (2025) 130390. 
\bibitem{k4} Kuate, P.D.K., Ito, H., Fossi, J.T. et al. From connectome to silicon: a biologically-inspired complex network of CMOS chaotic oscillators for analog brain emulation. Nonlinear Dyn (2025).
\bibitem{k5} Ishaq, M., Ashraf, M.B. Development of a novel artificial neural network-based approach for predicting entropy generation in electroosmotic flow of nanofluids. Nonlinear Dyn (2025).
\bibitem{k6} Drukarch B, Wilhelmus M M M and Shrivastava S 2021 The thermodynamic theory of action potential propagation: a sound basis for unification of the physics of nerve impulses Reviews in the Neurosciences 33 285302.
\bibitem{k7} Matan Mussel, Matthias F. Schneider. Sound pulses in lipid membranes and their potential function in biology. Progress in Biophysics and Molecular Biology 162 (2021) 101e110. 
\bibitem{k8} Heimburg T and Jackson A D 2005 On soliton propagation in biomembranes and nerves Proceedings of the National Academy of Sciences 102 97909795.  
\bibitem{k10} Rani, A.; Shakeel, M.; Kbiri Alaoui, M.; Zidan, A.M.; Shah, N.A.; Junsawang, P. Application of the $Exp-\varphi \xi$-Expansion Method to Find the Soliton Solutions in Biomembranes and Nerves. Mathematics 2022, 10, 3372. 
\bibitem{k11} El-Nabulsi RA. 2022 Emergence of lump-like solitonic waves in Heimburg–Jackson biomembranes and nerves fractal model. J. R. Soc. Interface 19: 20220079. 
\bibitem{k12} Razzaq, W., Akbulut, A., Zafar, A. et al. Solitary wave solutions of coupled nerve fibers model based on two analytical techniques. Opt Quant Electron 55, 591 (2023).
\bibitem{k13} González-Gaxiola, O.; Biswas, A.; Moraru, L.; Alghamdi, A.A. Solitons in Neurosciences by the Laplace–Adomian Decomposition Scheme. Mathematics 2023, 11, 1080.
\bibitem{k14} Shahzad T, Baber M Z, Qasim M, Sulaiman T A, Yasin M W and Ahmed N 2024 Explicit solitary wave profiles and stability analysis of biomembranes and nerves Modern Physics Letters B 38. 
\bibitem{k15} Tahira Jamal, Adil Jhangeer, Malik Zawwar Hussain. An anatomization of pulse solitons of nerve impulse model via phase portraits, chaos and sensitivity analysis. Chinese Journal of Physics 87 (2024) 496–509.
\bibitem{k16} Ozsahin D U, Ceesay B, baber M Z, Ahmed N, Raza A, Rafiq M, Ahmad H, Awwad F A and Ismail E A A 2024 Multiwaves, breathers, lump and other solutions for the heimburg model in biomembranes and nerves Scientific Reports 14.
\bibitem{k17} Younas, U., Muhammad, J., Almutairi, D.K. et al. Analyzing the neural wave structures in the field of neuroscience. Sci Rep 15, 7181 (2025).
\bibitem{k19} Aly R. Seadawy, Asghar Ali, Ahmet Bekir. Solitary wave solutions of the nonlinear fractional soliton neuron model via application of five mathematical methods.  Modern Physics Letters B (2025) 2550098 (20 pages). 
\bibitem{k9} C.S. Fedosejevs, \& M.F. Schneider, Sharp, localized phase transitions in single neuronal cells, Proc. Natl. Acad. Sci. U.S.A. 119 (8) e2117521119 (2022).
\bibitem{k20} Jüri Engelbrecht, Kert Tamm, Tanel Peets. On mathematical modelling of solitary pulses in cylindrical biomembranes. Biomech Model Mechanobiol (2015) 14:159–167. 
\bibitem{k21} Tanel Peets, Kert Tamm, Jüri Engelbrecht. On the role of nonlinearities in the Boussinesq-type wave equations. Wave Motion 71 (2017) 113–119. 
\bibitem{k22} Jüri Engelbrecht, Kert Tamm \& Tanel Peets (2017) On solutions of a Boussinesq-type equation with displacement-dependent nonlinearities: the case of biomembranes, Philosophical Magazine, 97:12, 967-987. 
\bibitem{k23} Tanel Peets, Kert Tamm, Päivo Simson, Jüri Engelbrecht. On solutions of a Boussinesq-type equation with displacement-dependent nonlinearity: A soliton doublet. Wave Motion 85(2019) 10–17. 
\bibitem{k18} J. A. Onana Inouga, S. E. Mkam Tchouobiap, M. Siewe Siewe, and F. M. Moukam Kakmeni. Action potential-like modes as modulated waves in an extended soliton model for biomembranes and nerves. AIP Advances 15, 015035 (2025). 
\bibitem{k24} Pavón-Torres, M A Agüero-Granados and R Valencia-Torres. Adiabatic evolution of solitons embedded in lipid membranes. Phys. Scr. 99 125256 (2024).
\bibitem{k25} H. C. Rosu and O. Cornejo-Pérez. Supersymmetric pairing of kinks for polynomial nonlinearities. Phys. Rev. E 71, 046607 (2005).
\bibitem{k26} O. Cornejo-Pérez and H. C. Rosu. Nonlinear Second Order Ode’s -factorization and particular solutions- Progress of Theoretical Physics, Vol. 114, No. 3 (2005).
\bibitem{k27} González, H.C. Rosu, O. Cornejo-Pérez, S.C. Mancas, Factorization conditions for nonlinear second-order differential equations, in: S. Manukure, W.-X. Ma (Eds.), Nonlinear and Modern Mathematical Physics-Proceedings 2022, Springer, USA, 2024, pp. 81–99. 
\bibitem{k28} G. González Contreras, Parametric factorization of nonlinear second order differential equations, Phys. Scr. 99 (2024) 055214.
\bibitem{k29} O. Cornejo-Pérez, P. Albares, J. Negro. Solutions of an extended Duffing–van der Pol equation with variable coefficients. Physica D 476 (2025), 134675. 
\bibitem{k32} Lee, JI., Werginz, P., Kameneva, T. et al. Membrane depolarization mediates both the inhibition of neural activity and cell-type-differences in response to high-frequency stimulation. Commun Biol 7, 734 (2024).
\bibitem{kk32} Pav\'on-Torres O., Agüero-Granados M. A. and Maguiña-Palma M. E. 2024 Interaction and adiabatic evolution of orthodromic and antidromic impulses in the axoplasmic fluid Physics Letters A 521 129740.
\bibitem{kkk32} Drab M, Daniel M, Kralj-Iglic V and Iglic A 2022 Solitons in the heimburgjackson model of sound propagation in lipid bilayers are enabled by dispersion of a stiff membrane The European Physical Journal E 45.
\bibitem{kk34} H. Nymeyer and H.-X. Zhou, A method to determine dielectric constants in nonhomogeneous systems: Application to biological membranes, Biophys. J. 94, 1185 (2008).
\bibitem{kk35} Row, Hyeongjoo and Fernandes, Joshua B. and Mandadapu, Kranthi K. and Shekhar, Karthik. Spatiotemporal dynamics of ionic reorganization near biological membrane interfaces. PHYSICAL REVIEW RESEARCH7,013185(2025).
\bibitem{k030} B. Lautrup, R. Appali, A.D. Jackson, T. Heimburg, The stability of solitons in biomembranes and nerves, Eur. Phys. J. E. Soft Matter 34 (6) (2011) 1–9.
\bibitem{k031} M.I. Perez-Camacho, J. Ruiz-Suarez, Propagation of a thermo-mechanical perturbation on a lipid membrane, Soft Matter 13 (2017) 6555–6561.
\bibitem{k032} H.F. Freistühler, J.H. Höwing, An analytical proof for the stability of Heimburg-Jackson pulses, 2013, arXiv:1303.5941 (math.AP).
\bibitem{kk30} Contreras F., Ongay F., Pav\'on O. and Aguero M. 2013 Non-topological solitons as travelling pulses along the nerve International Journal of Modern Nonlinear Theory and Application 02 195200.
\bibitem{k30} Abdelhalim Ebaid, Emad H. Aly. Exact solutions for the transformed reduced Ostrovsky equation via the F -expansion method in terms of Weierstrass-elliptic and Jacobian-elliptic functions. Wave Motion 49 (2012) 296-308. 
\bibitem{k31} Fongang Achu G, Mkam Tchouobiap S E, Moukam Kakmeni F M and Tchawoua C 2018 Periodic soliton trains and informational code structures in an improved soliton model for biomembranes and nerves Physical Review E 98.
\bibitem{k33} Nisha, Neetu Maan, Amit Goyal, Thokala Soloman Raju, C.N. Kumar. Chirped Lambert W-kink solitons of the complex cubic-quintic Ginzburg-Landau equation with intrapulse Raman scattering. Physics Letters A 384 (2020) 126675. 
\bibitem{k34} Amado, A., Mohammadi, A. A $\phi ^{6}$ soliton with a long-range tail. Eur. Phys. J. C 80, 576 (2020).
\bibitem{k35} Pavón-Torres O, Aguero M A, Belyaeva T L, Ramirez A and Serkin V N 2019 Unusual self-spreading or self-compression of the cubic-quintic nlse solitons owing to amplification or absorption Optik 184 446456.
\bibitem{k36} Pavón-Torres, O., Collantes-Collantes, J.R. \& Agüero-Granados, M.A. Quasi-stationary Evolution of Cubic-quintic NLSE Drop-like Solitons in DNA-protein Systems. Int J Theor Phys 64, 88 (2025).
\bibitem{k37} Wang M.L., Li X., Zhang J., The $G'/G$ - expansion method and evolution
erquations in mathematical physics. Phys. Lett. A, 372 (2008) 417 – 421.
\bibitem{k38} Kudryashov N.A., Seven common errors in finding exact solutions of nonlinear differential equations, Commun Nonlinear Sci Numer Simulat. 14 (2009) 3507 – 3529.
\bibitem{k39} Nikolai A. Kudryashov. A note on the $G'/G$ method. Applied Mathematics and Computation 217 (2010) 1755–1758. 
\end{thebibliography}
\end{document}